

\magnification=\magstep1
\baselineskip=15pt
\overfullrule=0pt
\nopagenumbers
\font\twelvebf=cmbx12

\rightline{hep-th/9210029}
\rightline{IASSNS-HEP-92/57}
\rightline{Aug 1992}
\vskip .7in
\centerline{\twelvebf The Nonlinear Schr\"odinger Equation and}
\centerline{\twelvebf Conserved Quantities in the}
\centerline{\twelvebf Deformed Parafermion and SL(2,R)/U(1) Coset Models}
\vskip 1in
\centerline{ Jeremy Schiff}
\smallskip
\centerline{\it Institute For Advanced Study}
\centerline{\it Olden Lane, Princeton, NJ 08540}
\vskip 1in
\centerline{\bf Abstract}
\smallskip
\noindent The relationship between the nonlinear Schr\"odinger hierarchy
and the parafermion and SL(2,R)/U(1) coset models, analogous to the
relationship between the KdV hierarchy and the minimal models, is explained.
To do this I first present an in depth study of a series of integrable
hierarchies related to NLS, and write down an action from which any of these
hierarchies, and the associated second Poisson bracket structures, can be
obtained. In quantizing the free part of this action we find many features
in common with the bosonized parafermion and SL(2,R)/U(1) models, and
particularly it is clear that the quantum NLS hamiltonians are conserved
quantities in these models. The first few quantum NLS hamiltonians are
constructed.

\vfill\eject

\footline={\hss\tenrm\folio\hss}
\pageno=1

\noindent{\bf 1. Introduction}

\noindent It is often stated that there are deep connections between
conformal field theories and integrable partial differential equations
in $1+1$ dimensions. The most cited pieces of evidence for this are
(1) that the second Poisson bracket structures associated with equations
of KdV type are classical limits of the conformal algebras [1] and (2) that
the integrals of motion that exist in certain perturbed conformal field
theories are quantum analogs of the conserved quantities (``hamiltonians'')
of certain integrable PDEs [2]. As has been appreciated in some of the
literature (see for example [3]), these two statements are really the same
in origin. Most conformal field theories studied to date are themselves
integrable, in the sense that in the enveloping algebra of the chiral algebra
there exist an infinite number of algebraically independent commuting
quantities, and that in some sense which we cannot currently define well,
the number of these ``integrals of motion '' is half the total number
of degrees of freedom of the system\footnote*{I thank E.Witten for persuading
me that generic conformal field theories are probably {\it not}
integrable. In any conformal field theory there will be an infinite number
of integrals of motion, just by virtue of the fact that the enveloping
algebra of the chiral algebra contains the Virasoro algebra. The question
is whether there are a sufficiently large number of them.}. Given an
integrable conformal field theory we can perturb by adding a hamiltonian
which is a sum of the conserved quantities; this will give a new theory
which is still integrable, having the same conserved quantities. In the
context of quantum field theory we are only really interested in relevant,
Lorentz invariant perturbations, i.e. adding to the action
(\`a la Zamolodchikov [2]) integrals of primary fields
with dimension less than two. This is in fact mathematically convenient;
it seems (but has certainly not been proven) that
in general in the conformal field theory there is not a unique way to choose
the conserved quantities, but the requirement that the set we choose
should include a particular physical hamiltonian pins down the set uniquely.
Assuming now that the chiral algebra has some identifiable classical
limit, it is quite clear that associated with any integrable conformal field
theory and choice of the set of conserved quantities there will be a
classical hierarchy of KdV type, with Poisson bracket algebra the classical
limit of the chiral algebra, and
with hamiltonians the classical limits of the quantum hamiltonians.

What might seem surprising is that this construction also seems to work in
reverse, that is, {\it all} known classical hierarchies of KdV type seem to
arise from the classical limit of some conformal field theory. To
understand this we need a
procedure to obtain a conformal field theory from a classical
hierarchy (from the Poisson bracket algebra of the hierarchy we can
presumably guess what the chiral algebra of the corresponding theory
is, but this is not sufficient). In [4] I described how to construct
an action that gave (for the correct choice of parameters)
an arbitrary equation in the KdV hierarchy as equation of
motion\footnote{**}{Since writing [4] I have been informed that this action
is actually well-known, see for example [5]; but in the literature I only
see it used as an action for the MKdV hierarchy, and in [4] crucial
information is gained by understanding how it serves as an action for
the KdV hierarchy too.}; this action consisted of a ``free'' part  and a
sum of terms proportional to the KdV hamiltonians.
I then showed that in quantizing the theory described by the ``free''
action, with a suitable choice of the coupling constant, we are led
to the Feigin-Fuchs description of the (holomorphic sector of the)
minimal models, and it becomes clear why the quantum KdV
hamiltonians are the conserved quantities associated
with the $\Phi_{(1,3)}$ perturbation of minimal models [2].
It is obviously of interest
to find generalizations of [4], that is to find actions for other
integrable hierarchies, and thereby hopefully reveal the underlying
conformal field theories; this paper is devoted to doing just this for
the case of the non-linear Schr\"odinger (NLS) hierarchy. We will see that
the NLS hierarchy is related to the parafermion and $SL(2,{\bf R})/U(1)$
coset models, and that the quantum NLS hamiltonians give us give
conserved quantities of these models and their deformations. Our classical
insight will give us a reasonably simple method for constructing
these charges.

The action construction for the KdV hierarchy exploited the
``quasi-hamiltonian'' (or ``antiplectic'') formalism for the KdV
hierarchy studied by Wilson [6] (see also [7]), and we need an
analog of this for the NLS hierarchy. The bulk of this paper (sections
2 and 3a) is devoted to an understanding of the NLS hierarchy and related
hierarchies, including the {\it two} ``Ur-NLS'' hierarchies. We establish
that the NLS hierarchy is just one in a large tower of hierarchies related
by Miura maps; the first non-trivial equations of the {\it ten} hierarchies
we will have reason to consider, and the principal Miura maps between them,
are given in figure 1. Any equation in any of these hierarchies can be
obtained by varying a single action with respect to appropriate fields.
I hope that the developments in sections 2 and 3a will be of interest
beyond the current conformal field theoretic application, and no knowledge
of conformal field theory is necessary to read them. I note that certain
hierarchies in the NLS tower have been of some interest recently in the context
of matrix models [8].

The detailed contents of sections 2 and 3a are as follows:
in 2a I introduce the NLS hierarchy and its hamiltonian structures;
in 2b I examine four related hierarchies essentially obtained from the
standard gauge equivalence class of the NLS hierarchy; in 2c I present
the two Ur-NLS hierarchies; in 2d I display the diverse symmetries of
the Ur-NLS quasi-hamiltonian structures, and use them to gain some insight
into the numerous hierarchies in the NLS tower; in 2e I discuss the
Toda flows associated with NLS; finally in 3a I
introduce three more hierarchies
and give the NLS action. The reader is advised to keep figure 1
handy throughout these sections.

The remainder of this paper is devoted to finding the conformal field
theory related to NLS, and exploiting our understanding of the classical
limit of this conformal field theory. In section 3b
I explain how the free part of the NLS action
is actually related to {\it two} different conformal field theories;
one of these is of little interest for reasons we shall see,
but the other is (depending on the value of the coupling constant) either a
parafermion or a $SL(2,{\bf R})/U(1)$ coset  model, and  I show that the
quantum NLS hamiltonians are conserved quantities of suitable deformations
of these models. In section 3c I point out how the free part of the NLS
action is apparently related to a gauged WZW model. This should come as no
surprise. Most, if not all, known integrable hierarchies are of
Drinfeld-Sokolov type [9] or its generalizations [10]. The second Poisson
bracket structures of these hierarchies are all reductions of a Kac-Moody
algebra and hence we can construct an associated gauged WZW model as a
candidate associated conformal field theory.
Finally in section 3d I construct the
first few quantum NLS hamiltonians; the results we find are in accord
with both the work of Fateev on deformed parafermionic models [11] and the work
of Bakas and Kiritsis on conserved quantities in the $SL(2,{\bf R})/U(1)$
coset models [12].
Despite work in the literature on both the classical limit of the nonlocal
parafermion algebra (which is the second hamiltonian structure of NLS)
and integrable deformations of the parafermion models, the relationship
of parafermions with NLS seems not to have been observed before.

One final comment is in order in this introduction.
In [13] a quantization of the NLS hierarchy was proposed
along the lines of the quantizations of the MKdV and KdV hierarchies
of Sasaki and Yamanaka [14]. Unfortunately in this work the {\it first}
Poisson bracket of the NLS hierarchy was quantized, and so it is not
directly relevant for our work in section 3d where we need quantum NLS
hamiltonians with respect to the {\it second} Poisson bracket (from work on
the quantum KdV equation we do not expect the bihamiltonian structure
to survive quantization). But the success of [13] is definitely a positive
sign for us.

\vskip.2in

\noindent{\bf 2. Understanding the Hierarchies related to NLS}

\noindent Throughout this paper I will use the title ``the NLS equation''
to refer to the coupled system of equations
$$\eqalign{\psi_t&=\psi_{xx}-2\psi^2\bar{\psi}\cr
           \bar{\psi}_t&=-\bar{\psi}_{xx}+2\bar{\psi}^2\psi} \eqno{(2.1)}$$
Here $\psi,\bar{\psi}$ are independent real fields and $x$ and $t$ are real
coordinates. The more usual form of the nonlinear Schr\"odinger equation is
obtained by considering (2.1) for $t$ pure imaginary, $\psi$ complex, and
$\bar{\psi}$ the complex conjugate of $\psi$ (then the two equations in (2.1)
become complex conjugates of each other).

\vskip.2in

\noindent {\it 2a. The bihamiltonian structure of (2.1) and the NLS hierarchy}

\noindent Magri [15] showed that (2.1) has a bihamiltonian structure. The first
hamiltonian structure is local, with Poisson brackets
$$ \pmatrix{\{{\psi}(x),{\psi}(y)\}&
            \{{\psi}(x),\bar{\psi}(y)\}\cr
            \{\bar{\psi}(x),{\psi}(y)\}&
            \{\bar{\psi}(x),\bar{\psi}(y)\}\cr}
  = \pmatrix{0&-1\cr 1&0}\delta(x-y) \eqno{(2.2)}$$
and hamiltonian $H_4=\int dx~(\psi_x\bar{\psi}_x+\psi^2\bar{\psi}^2)$. The
second hamiltonian structure is non-local, with brackets
$$ \pmatrix{\{{\psi}(x),{\psi}(y)\}&
            \{{\psi}(x),\bar{\psi}(y)\}\cr
            \{\bar{\psi}(x),{\psi}(y)\}&
            \{\bar{\psi}(x),\bar{\psi}(y)\}\cr}
  = \pmatrix{2{\psi}\partial_x^{-1}{\psi}&
             \partial_x - 2{\psi}\partial_x^{-1}\bar{\psi}\cr
             \partial_x - 2\bar{\psi}\partial_x^{-1}{\psi}&
             2\bar{\psi}\partial_x^{-1}\bar{\psi}\cr} \delta(x-y)
                           \eqno{(2.3)}$$
and hamiltonian $H_3=\int dx~\bar{\psi}\psi_x$. In (2.3) the matrix on the
RHS is evaluated at $x$, and the operator $\partial_x^{-1}$ are defined by
$\partial_x^{-1}f(x)={1 \over 2}(\int_0^x dx'f(x') + \int_{2\pi}^x dx'f(x'))$,
so that $\int_0^{2\pi} f_1(x)\partial_x^{-1}f_2(x) =
-\int_0^{2\pi} f_2(x)\partial_x^{-1}f_1(x)$ (the range of $x$ throughout this
paper is taken to be $[0,2\pi]$). An infinite number of conserved quantities
for the NLS equation are defined by the recursion relation
$$ \pmatrix{{{\delta H_{n+1}}\over{\delta\psi}}\cr
           {{\delta H_{n+1}}\over{\delta\bar{\psi}}}\cr} =
    \pmatrix{\partial_x - 2\bar{\psi}\partial_x^{-1}{\psi}&
             2\bar{\psi}\partial_x^{-1}\bar{\psi}\cr
             -2{\psi}\partial_x^{-1}{\psi}&
             -\partial_x + 2{\psi}\partial_x^{-1}\bar{\psi}\cr}
     \pmatrix{{{\delta H_n}\over{\delta\psi}}\cr
           {{\delta H_n}\over{\delta\bar{\psi}}}\cr}
                           \eqno{(2.4)}$$
The recursion is started by setting $H_2=-\int dx~\psi\bar{\psi}$, which
generates $H_3$,$H_4$ as given above. The infinite number of conserved
quantities generate higher equations in the NLS hierarchy, in the usual
fashion.

An important point about the Poisson bracket algebra (2.3), which underlies
much of what we will do later, was pointed out in [16]; the
algebra (2.3) is precisely the Dirac reduction [17] of the $SL(2)$ Kac-Moody
algebra under the constraint that the diagonal current is set to zero.

\vskip.2in

\noindent{\it 2b. The gauge equivalence class of the NLS equation}

\noindent The NLS equation is a representative of a whole class of
integrable evolution equations whose general form is
$$ \eqalign{
  d_t&=-2\phi_x-(ef)_x\cr
  e_t&=e_{xx}+4e(\phi+d^2)+2ed_x+4de_x\cr
  f_t&=-f_{xx}-4f(\phi+d^2)+2fd_x+4df_x  } \eqno{(2.5)}$$
This set of equations emerges naturally in the context of reductions of
the self-dual Yang-Mills equations; see for example [18].
Equations (2.5) describe the evolutions of functions $d,e,f$ in terms of
$d,e,f$ and some auxiliary field $\phi$. The presence of $\phi$ allows us
to choose some combination of $d,e,f$ that we set to zero (this is called
a ``gauge choice''), only we then have to adjust $\phi$ so that the
time derivative of our constraint is zero. This freedom arises because
for an arbitrary function $\alpha$, the replacements
$$\eqalign{d&\rightarrow d+\alpha_x/\alpha\cr
           e&\rightarrow e\alpha^{-2}\cr
           f&\rightarrow f\alpha^2\cr
           \phi&\rightarrow \phi-\alpha_t/2\alpha}\eqno{(2.6)}$$
map solutions of (2.5) to solutions. The NLS equation (2.1) is
(2.5) with $d=0$, $\phi=-{1 \over 2}ef$ (and $e=\psi$, $f=\bar{\psi}$).
Note that setting $d=0$ in (2.5) does not uniquely determine $\phi$, so this
is an {\it incomplete} gauge fixing; looking at (2.6), we see that this
corresponds to the invariance of the (2.1) under $\psi\rightarrow\alpha^{-2}
\psi$, $\bar{\psi}\rightarrow\alpha^{2}\bar{\psi}$ where $\alpha$ is constant.
Possible {\it complete} gauge fixings include

\noindent 1. $e=1$, $d=A$, $f=B$ ($\phi=-d^2-{1\over 2}d_x$).

\noindent The equations are
$$ \eqalign{A_t&=A_{xx}+(2A^2-B)_x\cr
            B_t&=-B_{xx}+(4AB)_x} \eqno{(2.7)}$$
A solution of (2.1) gives a solution of (2.7) via
$$ \eqalign{A&=\psi_x/2\psi\cr
            B&=\psi\bar{\psi}} \eqno{(2.8)}$$

\noindent 2. $f=1$, $d=C$, $e=D$ ($\phi=-d^2+{1\over 2}d_x$).

\noindent The equations are
$$ \eqalign{C_t&=-C_{xx}+(2C^2-D)_x\cr
            D_t&=D_{xx}+(4CD)_x}\eqno{(2.9)}$$
A solution of (2.1) gives a solution of (2.9) via
$$ \eqalign{C&=-\bar{\psi}_x/2\bar{\psi}\cr
            D&=\psi\bar{\psi}}\eqno{(2.10)}$$

\noindent There is an obvious invertible
map from the solutions of (2.7) to those
of (2.9); set $C=-A$, $D=B$ and let $t\rightarrow -t$. But from the gauge
transformation relations we get another invertible
map not involving change of coordinates
$$ \eqalign{C&=A-B_x/2B\cr
            D&=B}\eqno{(2.11)}$$
Equations (2.7) and (2.9) are thus equivalent, and in the sequel
we will just use (2.7). The bihamiltonian structure of the NLS equation (2.1)
induces a {\it local} bihamiltonian structure for (2.7), the Poisson brackets
being
$$ \pmatrix{ \{A(x),A(y)\} &
             \{A(x),B(y)\} \cr
             \{B(x),A(y)\} &
             \{B(x),B(y)\} \cr}
   = -{\textstyle{1\over 2}}
                \pmatrix{0&\partial_x\cr\partial_x&0\cr}\delta(x-y)
                 \eqno{(2.12)}$$
$$ \pmatrix{ \{A(x),A(y)\} &
             \{A(x),B(y)\} \cr
             \{B(x),A(y)\} &
             \{B(x),B(y)\} \cr}
   = {\textstyle{1\over 2}}
        \pmatrix{-\partial_x&\partial_x(\partial_x+2A)\cr
                 -(\partial_x-2A)\partial_x&2(\partial_xB+B\partial_x)\cr}
                 \delta(x-y)
                 \eqno{(2.13)}$$
(induced by (2.2) and (2.3) respectively). The hamiltonian
is $H_4=2\int dx~(AB_x-2A^2B+{1\over 2}B^2)$ for the brackets (2.12)
and $H_3=2\int dx~AB$ for the brackets (2.13). The recursion relation
for the conserved quantities of equation (2.7) (which reduce
to those of the NLS equation via the substitution (2.8)) is
$$ \pmatrix{ {{\delta H_{n+1}}\over{\delta A}} \cr
             {{\delta H_{n+1}}\over{\delta B}}\cr }
  = \pmatrix{ \partial_x-2\partial_x^{-1}A\partial_x &
               -2( B+\partial_x^{-1}B\partial_x)\cr
               1&
               -(\partial_x+2A)\cr }
  \pmatrix{ {{\delta H_{n}}\over{\delta A}}\cr
             {{\delta H_{n}}\over{\delta B}}\cr } \eqno{(2.14)}$$
and $H_2=-\int dx~B$, $H_1=-\int dx~A$. The $A,B$ hierarchy is
an instance of the Broer-Kaup hierarchy studied in [19].

The relationship between equations (2.1) and (2.7) is somewhat analogous
to the relationship between the modified KdV and KdV equations. I
therefore propose for (2.7) the name the {\it unmodified} NLS
($\rm M^{-1}NLS$) equation.

One other incomplete gauge fixing of the system (2.5) is important.
Requiring $2d-e+f=0$, we can take $\phi=-{1\over 4}(e_x+f_x+(e-f)^2)$.
Writing $e=j$, $f=\bar{j}$ we obtain the equations
$$ \eqalign{j_t&=(j_x+j^2-2j\bar{j})_x\cr
            \bar{j}_t&=(-\bar{j}_x-\bar{j}^2+2j\bar{j})_x}\eqno{(2.15)}$$
These equations appeared in [20]. Since both (2.1) and (2.15) are
incomplete gauge fixings, there is no explicit Miura map between them,
but from the gauge transformation laws (2.6) it can be seen that
a solution of (2.1) can be found from a solution of (2.15) by
quadratures. Solutions of (2.15) generate solutions of (2.7)
directly, however, via the map
$$ \eqalign{A&={\textstyle{1\over 2}}(j-\bar{j}+j_x/j)\cr
            B&=j\bar{j}}\eqno{(2.16)}$$
For want of a better name I will call (2.15) the NLS2 equation. In
the ``tower'' of equations we are building around the NLS equation, the
NLS2 equation is in three ways more analogous to the MKdV equation than
the original NLS equation:

\noindent 1. Whereas for NLS the first bracket
structure was local and the second non-local, for NLS2 the first
bracket structure is non-local and the second local;
explicitly they are [20]
$$ \pmatrix{\{j(x),j(y)\}&
            \{j(x),\bar{j}(y)\}\cr
            \{\bar{j}(x),j(y)\}&
            \{\bar{j}(x),\bar{j}(y)\}\cr}=\hfill $$
$$ \pmatrix{j(\partial_x+j+\bar{j})^{-1}+(\partial_x-j-\bar{j})^{-1}j&
         -1+j(\partial_x+j+\bar{j})^{-1}-(\partial_x-j-\bar{j})^{-1}\bar{j}\cr
          1-\bar{j}(\partial_x+j+\bar{j})^{-1}+(\partial_x-j-\bar{j})^{-1}j&
     -\bar{j}(\partial_x+j+\bar{j})^{-1}-(\partial_x-j-\bar{j})^{-1}\bar{j}\cr}
             \delta(x-y)
           \eqno{(2.17a)}$$
$$ \pmatrix{\{j(x),j(y)\}&
            \{j(x),\bar{j}(y)\}\cr
            \{\bar{j}(x),j(y)\}&
            \{\bar{j}(x),\bar{j}(y)\}\cr}
  =  \pmatrix{0&\partial_x\cr\partial_x&0}
             \delta(x-y)
           \eqno{(2.17b)}$$

\noindent 2.
The right hand sides of (2.15) are total $x$-derivatives, unlike those of
(2.1), so from (2.15) we can naturally define a ``potential'' NLS equation;
introducing $h,\bar{h}$ by $j=h_x$, $\bar{j}=\bar{h}_x$ we see that if
$h,\bar{h}$ satisfy
$$ \eqalign{h_t&=h_{xx}+h_x^2-2h_x\bar{h}_x\cr
            \bar{h}_t&=-\bar{h}_{xx}-\bar{h}_x^2+2h_x\bar{h}_x}
                             \eqno{(2.18)}$$
then $j,\bar{j}$ satisfy (2.15). In fact solutions of (2.18) also give
rise to solutions of (2.1), but the Miura map is less obvious:
$$ \eqalign{\psi&=e^{h-\bar{h}}h_x\cr
            \bar{\psi}&=e^{-(h-\bar{h})}\bar{h}_x} \eqno{(2.19)}$$

\noindent 3. As noted above, the NLS equation has a very obvious
invariance under $\psi\rightarrow\alpha^{-2}\psi$,
$\bar{\psi}\rightarrow\alpha^{2}\bar{\psi}$, which is effectively
``modded out'' when we pass to the $\rm M^{-1}NLS$ equation. The
corresponding invariance of NLS2 is hidden, like that of the MKdV
equation [6].

For completeness I note that the recursion operator
for the NLS2 equation is
$$ \pmatrix{\partial_x+\partial_x^{-1}j_x+\bar{j}-j &
            \partial_x^{-1}\bar{j}_x-2\bar{j} \cr
            -\partial_x^{-1}j_x+2j &
            -\partial_x-\partial_x^{-1}\bar{j}_x+\bar{j}-j \cr}
       \eqno{(2.20)}$$
and the first few conserved quantities are
$$ \eqalign{H_1&= -{\textstyle{1\over 2}}\int dx~(j-\bar{j})  \cr
            H_2&=-\int dx~ j\bar{j}  \cr
            H_3&=\int dx~ (\bar{j}j_x+(j-\bar{j})j\bar{j}) } \eqno{(2.21)}$$
Note also that the NLS2 equation (2.15) reduces to Burger's equation [21]
on setting $\bar{j}=0$.

So far in this section
I have described the NLS, $\rm M^{-1}NLS$, and NLS2 equations,
all of which emerge from the scheme (2.5), and the potential
NLS equation, which has an ``obvious'' relationship to NLS2.
There is one more equation that has an ``obvious'' relationship to NLS2;
given a solution of NLS2, construct $U,V$ by
$$ \eqalign{ U&=j-\bar{j} \cr
             V&=(j+\bar{j})_x-{\textstyle{1\over 2}}(j+\bar{j})^2 }
                 \eqno{(2.22)}$$
Then $U,V$ satisfy
$$ \eqalign{ U_t&=V_x+3UU_x\cr
             V_t&=U_{xxx}+V_xU+2U_xV} \eqno{(2.23)}$$
I will call this the $\rm M^{-1}NLS2$ equation.
$\rm M^{-1}NLS2$ inherits {\it local} bihamiltonian structure
{}from NLS2, with brackets
$$ \pmatrix{ \{U(x),U(y)\} &
             \{U(x),V(y)\} \cr
             \{V(x),U(y)\} &
             \{V(x),V(y)\} \cr}
  = 2\pmatrix{0&\partial_x\cr\partial_x&-(\partial_xU+U\partial_x)\cr}
             \delta(x-y)
             \eqno{(2.24)}$$
$$ \pmatrix{ \{U(x),U(y)\} &
             \{U(x),V(y)\} \cr
             \{V(x),U(y)\} &
             \{V(x),V(y)\} \cr}
  =  -2\pmatrix{\partial_x&0\cr 0&\partial_x^3+V\partial_x+\partial_xV\cr}
             \delta(x-y)
             \eqno{(2.25)}$$
The respective hamiltonians are $H_4={1\over 4}\int dx~(V^2+3U^2V+{5\over 4}
U^4-U_x^2)$ and $H_3=-{1\over 2}\int dx~(UV+{1\over 2}U^3)$.
Further hamiltonians can be defined by the recursion relation
$$ \pmatrix{ {{\delta H_{n+1}}\over{\delta U}} \cr
             {{\delta H_{n+1}}\over{\delta V}}\cr }
  = -\pmatrix{ U+\partial_x^{-1}U\partial_x &
               \partial_x^2+V+\partial_x^{-1}V\partial_x\cr
               1&
               0\cr}
  \pmatrix{ {{\delta H_{n}}\over{\delta U}}\cr
             {{\delta H_{n}}\over{\delta V}}\cr } \eqno{(2.26)}$$
and $H_1=-{1\over 2}\int dx~U$, $H_2={1\over 2}\int dx~(V+{1\over 2}U^2)$.
It is not clear that the hamiltonians defined here coincide with those
we have used previously, but it turns out this is indeed the case (i.e.
$H_n[U,V]$ and $H_n[A,B]$ agree when written as functionals of $j,\bar{j}$).

Now in some sense $\rm M^{-1}NLS$ and $\rm M^{-1}NLS2$ are not
really distinct. If
$U,V$ satisfy $\rm M^{-1}NLS2$ (2.23) then
$$ \eqalign{\tilde{A}&={\textstyle{1\over 2}}U\cr
            \tilde{B}&={\textstyle{1\over 4}}(-2V+2U_x-U^2) }\eqno{(2.27)}$$
satisfy $\rm M^{-1}NLS$ (2.7). Furthermore (2.27) maps the second hamiltonian
structure (2.25) of $\rm M^{-1}NLS2$ to that (2.13) of $\rm M^{-1}NLS$, and
the first hamiltonian structure (2.24) is taken to {\it minus} (2.12).
So although because of this minus sign the map (2.27) is not strictly
a Miura map of the $\rm M^{-1}NLS2$ {\it hierarchy} to
the $\rm M^{-1}NLS$ {\it hierarchy}, we might wish to consider
$\rm M^{-1}NLS$ and $\rm M^{-1}NLS2$ as equivalent.
But remarkably the composite
Miura map from NLS2 to $\rm M^{-1}NLS$ defined by (2.22) and (2.27) is
{\it not} the same as the direct map (2.16). So if we were to regard
$\rm M^{-1}NLS$ and $\rm M^{-1}NLS2$ as the same, each time we wished to
use a Miura map to pass from NLS2 to the single ``unmodified'' equation, we
would have to specify which map we were using. This is potentially
confusing. So we will opt, in the rest of this paper,
to treat $\rm M^{-1}NLS$ and $\rm M^{-1}NLS2$ as
distinct; thus all our Miura maps are genuine maps of hierarchies,
and different Miura maps go to different equations. Still, the existence
of two distinct Miura maps from the NLS2 equation to the
$\rm M^{-1}NLS$ equation is a most remarkable result.

To summarize section 2b:
using the scheme (2.5) we have obtained five of the equations
indicated in figure 1. I should point out here that (2.5)
is really only a restricted set of the
equations in the gauge equivalence class of the NLS equation. For further
details see [18].

\vskip.2in

\noindent{\it 2c. Antiplectic formalism for the NLS equation}

\noindent From the work of Wilson [6] we recognize the need to find
an ``Ur-NLS'' hierarchy; this should generate solutions of the potential
NLS hierarchy (2.18) via a Miura map (and hence will give us solutions
of the NLS, NLS2, $\rm M^{-1}NLS$, and $\rm M^{-1}NLS2$ hierarchies),
and it should have an {\it inverse-local} bihamiltonian form.
We expect the two inverse-local hamiltonian structures to induce
the second and {\it third} Poisson bracket structures of the various equations
of section 2b. The third Poisson bracket structures are obtained by letting
the recursion operators act (from the right) on the matrices appearing
in the second bracket structures; I will just give the explicit
form of the third bracket structure for the NLS2 equation, which is of
particular interest since it is local [20]:
$$ \pmatrix{\{j(x),j(y)\}&
            \{j(x),\bar{j}(y)\}\cr
            \{\bar{j}(x),j(y)\}&
            \{\bar{j}(x),\bar{j}(y)\}\cr}
  =  \pmatrix{\partial_x j + j\partial_x&
              -\partial_x^2-\partial_x j + \bar{j}\partial_x\cr
              \partial_x^2+\partial_x\bar{j}-j\partial_x&
             -(\partial_x\bar{j}+\bar{j}\partial_x)\cr}
             \delta(x-y)
           \eqno{(2.28)}$$

It turns out that there are actually {\it two}
Ur-NLS hierarchies (related in the same way as $\rm M^{-1}NLS$ and
$\rm M^{-1}NLS2$).  I do not plan to give a full
derivation here of these hierarchies, since much will be obvious after
reading sections 2d and 3a below; here I will just give the equations,
their two Poisson bracket structures and the associated {\it local}
symplectic forms, and the Miura maps to the potential NLS hierarchy.

\vskip.1in

I will call the following equation the Ur-NLS1 equation:
$$ \eqalign{ T_t&=T_{xx}+2T_xS_x\cr
             S_t&={{2S_xT_{xx}}\over{T_x}}+3S_x^2-S_{xx}  } \eqno{(2.29)}$$
It is a straightforward but tedious business to check the following
results:

\noindent 1. If $S,T$ satisfy Ur-NLS1 then
$$ \eqalign{h&=S\cr
            \bar{h}&=-S+\ln(S_xT_x^{-1})  }
          \eqno{(2.30)}$$
satisfy the potential NLS equation.

\noindent 2. The bracket structure
$$ \pmatrix{ \{S(x),S(y)\} &
             \{S(x),T(y)\} \cr
             \{T(x),S(y)\} &
             \{T(x),T(y)\} \cr}
  = \pmatrix{ 0 &
              -\partial_x^{-1}T_x\partial_x^{-1} \cr
             \partial_x^{-1}T_x\partial_x^{-1} &
             2\partial_x^{-1}T_x\partial_x^{-1}T_x\partial_x^{-1} \cr}
      \delta(x-y) \eqno{(2.31)}$$
induces, via the various Miura maps given, the second hamiltonian structures
of the NLS, NLS2,  $\rm M^{-1}NLS$ and  $\rm M^{-1}NLS2$ equations.
The inverse of the matrix on the right hand side of (2.31) is
$$ \pmatrix{ 2\partial_x &
             \partial_xT_x^{-1}\partial_x \cr
             -\partial_xT_x^{-1}\partial_x &
             0 \cr}    \eqno{(2.32)}$$
which is a local operator, associated with the symplectic form
$$ \Omega_2=2\int dx~\delta(S+\ln T_x)\wedge\delta S_x    \eqno{(2.33)}$$

\noindent 3. The bracket structure
$$ \pmatrix{ \{S(x),S(y)\} &
             \{S(x),T(y)\} \cr
             \{T(x),S(y)\} &
             \{T(x),T(y)\} \cr}
  = -\pmatrix{S_x\partial_x^{-1}+\partial_x^{-1}S_x&
              \partial_x^{-1}T_x\cr
              T_x\partial_x^{-1}&
              0\cr}
      \delta(x-y) \eqno{(2.34)}$$
induces, via the various Miura maps given, the third hamiltonian structures
of the NLS, NLS2,  $\rm M^{-1}NLS$ and  $\rm M^{-1}NLS2$ equations.
The inverse of the matrix on the right hand side of (2.34) is
$$ \pmatrix{ 0&
             -\partial_xT_x^{-1}\cr
             -T_x^{-1}\partial_x&
             \partial_xS_xT_x^{-2}+S_xT_x^{-2}\partial_x\cr}    \eqno{(2.35)}$$
which is a local operator, associated with the symplectic form
$$ \Omega_3=-2\int dx~\delta T\wedge\delta\left({{S_x}\over{T_x}}\right)
                             \eqno{(2.36)}$$

\vskip.1in

I will call the following equation the Ur-NLS2 equation:
$$ \eqalign{ \tau_t&=-\tau_{xx}-2\tau_x\sigma_x\cr
             \sigma_t&=\sigma_{xx}-3\sigma_x^2-{{2\sigma_x\tau_{xx}}\over
                       {\tau_x}}  } \eqno{(2.37)}$$
This is (2.29) with $t\rightarrow -t$. We check:

\noindent 1. If $\sigma,\tau$ satisfy Ur-NLS2 then
$$ \eqalign{h&=-\sigma\cr
            \bar{h}&=\sigma+\ln\tau_x  }
          \eqno{(2.38)}$$
satisfy the potential NLS equation.

\noindent 2. The bracket structure
$$ \pmatrix{ \{\sigma(x),\sigma(y)\} &
             \{\sigma(x),\tau(y)  \} \cr
             \{\tau(x),\sigma(y)  \} &
             \{\tau(x),\tau(y)    \} \cr}
  = \pmatrix{ 0 &
             - \partial_x^{-1}\tau_x\partial_x^{-1} \cr
             \partial_x^{-1}\tau_x\partial_x^{-1} &
             2\partial_x^{-1}\tau_x\partial_x^{-1}\tau_x\partial_x^{-1} \cr}
      \delta(x-y) \eqno{(2.39)}$$
induces, via the various Miura maps given, the second hamiltonian structures
of the NLS, NLS2,  $\rm M^{-1}NLS$ and  $\rm M^{-1}NLS2$ equations.
Comparing with (2.31) we see that the inverse of the matrix on the right
hand side of (2.39) is a local operator, associated with the symplectic form
$$ \tilde{\Omega}_2=2\int dx~\delta(\sigma+\ln\tau_x)\wedge\delta\sigma_x
                 \eqno{(2.40)}$$

\noindent 3. The bracket structure
$$ \pmatrix{ \{\sigma(x),\sigma(y)\} &
             \{\sigma(x),\tau(y)  \} \cr
             \{\tau(x),\sigma(y)  \} &
             \{\tau(x),\tau(y)    \} \cr}
  =\pmatrix{\sigma_x\partial_x^{-1}+\partial_x^{-1}\sigma_x &
             \partial_x^{-1}\tau_x\cr
             \tau_x\partial_x^{-1}&
             0\cr}    \delta(x-y)            \eqno{(2.41)}$$
induces, via the various Miura maps given, the third hamiltonian structures
of the NLS, NLS2,  $\rm M^{-1}NLS$ and  $\rm M^{-1}NLS2$ equations.
Comparing with (2.34), the inverse of the matrix on the right hand side
of (2.41) is a local operator, associated with the symplectic form
$$ \tilde{\Omega}_3=2\int dx~\delta\tau\wedge\delta\left(
        {{\sigma_x}\over{\tau_x}}\right)    \eqno{(2.42)}$$

\vskip.1in

As mentioned above we see that the relationship between the two Ur-NLS
hierarchies is the same as that between the two $\rm M^{-1}NLS$ hierarchies.
The two Ur-NLS equations are related by a $t\rightarrow -t$
transformation, the second bracket structures are identical, and the third
bracket structures are identical up to an overall minus sign.
Again though, because of these sign differences, and more importantly because
of the different forms of the Miura maps (2.30) and (2.38), we will
regard Ur-NLS1 and Ur-NLS2 as distinct.
In sections 2d and  3 we will see how  Ur-NLS1 is related to
$\rm M^{-1}NLS$ and how Ur-NLS2 is related to $\rm M^{-1}NLS2$, thus
explaining why I have chosen to number the two equations in this way.

\vskip.2in

\noindent{\it 2d. Group-theoretical origin of the antiplectic formalism}

\noindent The symplectic forms $\Omega_2$, $\Omega_3$ that appeared
in section 2c in the context of the Ur-NLS1 equation,
have a natural group theoretical origin.
Consider an $SL(2)$ matrix valued function $g(x)$ such that the
diagonal component of $g^{-1}g_x$ vanishes.
Choosing a Gauss decomposition for $g$
$$ g=\pmatrix{1&T\cr0&1\cr}
     \pmatrix{e^{-S}&0\cr0&e^S\cr}
     \pmatrix{1&0\cr\phi&1\cr}  \eqno{(2.43)}$$
where $S,T,\phi$ are functions of $x$, the requirement that
the diagonal component of $g^{-1}g_x$ vanishes can be solved to give
$$ \phi={{S_xe^{-2S}}\over{T_x}} \eqno{(2.44)}$$
and we then have
$$ g = \pmatrix{ e^{-S}(1+TS_xT_x^{-1}) &
                 Te^S\cr
                 S_xT_x^{-1}e^{-S}&
                 e^S\cr}    \eqno{(2.45)}$$
With this choice of $g$ it can be checked that
$$\eqalign{
  \Omega_2&=-\int dx~ Tr[\delta g g^{-1}\wedge\partial_x(\delta g g^{-1})]\cr
  \Omega_3&={1\over 2}\int dx~ Tr\left[g^{-1}\delta g \wedge \left[
         \pmatrix{1&0\cr0&-1\cr},g^{-1}\delta g\right]\right] }\eqno{(2.46)}$$
The first of these formulas is an analogue of the one given for the KdV
equation in [6]. For the KdV equation the SL(2) matrix $g$ has to
satisfy the constraint that the (12) entry of $g^{-1}g_x$ is 1; this
constraint can be solved to give all the components of $g$ in terms of
two functions, but it turns out that the form $\Omega_2$ is independent
of one of these. The resulting symplectic form is the form associated with
the second hamiltonian structure of the Ur-KdV equation. The symplectic
form associated with the third hamiltonian structure of the Ur-KdV
equation can also be obtained from a formula similar to the second one
in (2.46); just the matrix appearing should be replaced by
$\pmatrix{0&0\cr1&0\cr}$ [18].

Having obtained the formula for $\Omega_2$ in (2.46) we can now
examine its symmetries, to obtain an ``explanation'', in the sense of
Wilson, as to why some of the equations in the NLS tower exist. Looking
at (2.33) we observe that $\Omega_2$ is invariant under $S\rightarrow S+s$,
$T\rightarrow T+t$, where $s,t$ are constants, but from (2.46) we see
that it is also invariant under $g\rightarrow hg$ where $h$ is a constant
$SL(2)$ matrix (and this preserves the constraint that $g^{-1}g_x$ should
have vanishing diagonal component). Writing
$$ h=\pmatrix{a&b\cr c&d\cr} \eqno{(2.47)}$$
where $ad-bc=1$, the corresponding action on $S,T$ is given by
$$ \eqalign{ e^S &\rightarrow (cT+d)e^S \cr
             T   &\rightarrow {{aT+b}\over{cT+d}} } \eqno{(2.48)}$$
Taking $a=d=1$, $c=0$ we replicate the translational symmetry in $T$
observed before; the translational symmetry in $S$ is however distinct from
the $SL(2)$ invariance. Thus we have in total an $SL(2)\times{\bf R}$ symmetry
group, with parameters $a,b,c,d,s$ with $ad-bc=1$.
Some of the equations in the NLS
tower now become clear. In moving from Ur-NLS1 to potential NLS we
mod out by just the $T$ translational symmetry; in moving from Ur-NLS1
to NLS we mod out by the full $SL(2)$ symmetry, leaving the non-hidden
residual one-parameter symmetry that we have mentioned; in moving from
Ur-NLS1 to NLS2 however, we mod out by the $S$ translational symmetry and
the subgroup of upper triangular matrices in $SL(2)$, which is not a normal
subgroup and hence leaves us with a hidden residual one-parameter symmetry;
finally in moving to the $\rm M^{-1}NLS$ equations we mod out by the full
$SL(2)\times{\bf R}$ symmetry.

But what of the Ur-NLS2 and $\rm M^{-1}NLS2$ equations?
Because $\Omega_2$ and $\tilde{\Omega}_2$ are identical we know that
$\tilde{\Omega}_2$ is invariant under translation of $\sigma$ and an
$SL(2)$ action like (2.48). Comparing (2.30) and (2.38) we see that
$S$ translation and $\sigma$ translation are equivalent, but the $SL(2)$
actions on the $S,T$ variables and the $\sigma,\tau$ variables are at least
partially distinct (by this I mean that their induced actions on the
$h,\bar{h}$ variables are distinct since we cannot  compare their full
actions directly). One way to verify that the two $SL(2)$ actions
are distinct is to check that under the $S,T$ $SL(2)$
action $A,B$ are invariant but $U,V$ are not, while under the
$\sigma,\tau$ $SL(2)$ action $U,V$ are invariant, but $A,B$ are not.
So we are led to the
conclusion that the NLS tower really consists of two sub-towers, the first
($\rm Ur-NLS1\rightarrow PotNLS\rightarrow NLS~or~NLS2 \rightarrow M^{-1}NLS$)
related to the symmetry in the $S,T$ variables, and the second
($\rm Ur-NLS2\rightarrow PotNLS\rightarrow NLS2 \rightarrow M^{-1}NLS2$)
related to the symmetry in the $\sigma,\tau$ variables. That these two
sub-towers have equations in  common may be a coincidence, or may
indicate some higher structure, namely the existence of a ``$\rm M^{-2}NLS$''
equation for variables which are invariant under {\it both} $SL(2)$ actions,
and possibly a corresponding ``$\rm Ur^2-NLS$'' equation for variables
upon which both $SL(2)$ actions are manifest. I have been unable to find
such a structure to date.


\vskip.2in

\noindent{\it 2e. Additional Flows}

\noindent To complete the picture we have of the NLS system, we need one
more ingredient. The NLS hamiltonians
form a ``complete set'', in the sense that there are no more
functionals of $A,B$ (or $U,V$)
which are integrals of local densities and commute
with the $H_n$, $n=1,2,3,...$ given above [13]. However if we are willing to
consider integrals of local densities in $h,\bar{h}$ then there are some
additional hamiltonians we should look at;  these generate flows that
are the analogs of the Liouville and Sine-Gordon flows in KdV theory.
Introduce
$$\eqalign{{\cal H}_1~&=~\int dx~e^{h+\bar{h}} \cr
           {\cal H}_2~&=~\int dx~h_xe^{-(h+\bar{h})} } \eqno{(2.49)}$$
${\cal H}_1$,${\cal H}_2$ have simple interpretations. From section 2d
there are two one-parameter hidden symmetries of the $j,\bar{j}$ system
(the residual symmetries from the two $SL(2)$ actions that we have not yet
removed). ${\cal H}_1$ and ${\cal H}_2$ are the generators of these
symmetries (${\cal H}_1$ comes form the $\sigma,\tau$ $SL(2)$ action,
and ${\cal H}_2$ from the $S,T$ $SL(2)$ action). Hence it is evident
that ${\cal H}_1$ commutes with $H_n$ when written as a functional of
$U,V$, since it commutes with $U$ and $V$, and it is evident
that ${\cal H}_2$ commutes with $H_n$ when written as a functional of
$A,B$, since it commutes with $A$ and $B$. Note that ${\cal H}_1$
vanishes when we write it in terms of $\sigma,\tau$, and assume periodic
boundary conditions,  and ${\cal H}_2$ similarly
vanishes when we write it in terms of $S,T$. Note that ${\cal H}_1$
and ${\cal H}_2$ do not commute with each other.

The actual forms of the flows will interest us little in the sequel, but
for completeness I will write them down; a hamiltonian $\mu_1{\cal H}_1+
\mu_2{\cal H}_2$ generates the flow
$$ \eqalign{ h_{xt}      &=\mu_1e^{h+\bar{h}} -\mu_2h_xe^{-(h+\bar{h})} \cr
             \bar{h}_{xt}&=\mu_1e^{h+\bar{h}} +\mu_2\bar{h}_xe^{-(h+\bar{h})}
    }\eqno{(2.50)}$$
We see that ${\cal H}_1$ generates a Liouville flow (for $h+\bar{h}$), but
${\cal H}_2$ generates an interesting integrable flow that I have not seen
before in the literature.

\vskip.2in

\noindent{\bf 3. The NLS Action and Associated Quantum Theories}

\noindent{\it 3a. The NLS action}

\noindent To readers of [4] it will come as no surprise that {\it every
equation in the NLS tower can be derived from a single action by varying
it with respect to different fields}. The action can be written
$$ S_{NLS}~=~\tilde{k}
             \int dxdt~h_x\bar{h}_t~+~\sum_{n=1}^{\infty}\lambda_n\int dt~H_n
                ~+~\sum_{i=1}^2\mu_i\int dt~{\cal H}_i  \eqno{(3.1)}$$
Here $\tilde{k},
\lambda_n,\mu_i$, $n=1,2,...$, $i=1,2$, are constants. In particular:
the $\rm M^{-1}NLS$ hierarchy is obtained by varying $S_{NLS}$ (with $\mu_i=0$)
with respect to the Ur-NLS1 variables $S,T$ and vice-versa;
the $\rm M^{-1}NLS2$ hierarchy is obtained by varying $S_{NLS}$
(with $\mu_i=0$) with respect to the Ur-NLS2 variables $\sigma,\tau$
and vice-versa; the NLS2 hierarchy and the ``additional flows'' (2.50)
are obtained by varying $S_{NLS}$ with respect to the PotNLS1 variables
$h,\bar{h}$;
the PotNLS hierarchy is obtained by varying $S_{NLS}$ (with $\mu_i=0$)
with respect to the NLS2 variables $j,\bar{j}$.
To obtain the original NLS hierarchy from (3.1) we need to introduce a
new set of variables $p=T$, $q=S_x/T_x$; if $S,T$ satisfy the Ur-NLS1
equation then $p,q$ satisfy what we shall call the AuxNLS equation,
$$ \eqalign{ p_t&=p_{xx}+2p_x^2q \cr
             q_t&=-q_{xx}+(2q^2p_x)_x }\eqno{(3.2)}$$
A Miura map from the AuxNLS equation to the NLS2 equation is given by
$$ \eqalign{ j&=qp_x\cr
             \bar{j}&=-qp_x+q_x/q }\eqno{(3.3)}$$
We can exploit this to vary $S_{NLS}$ (with $\mu_i=0$)
with respect to $p,q$ to obtain the NLS hierarchy; similarly we obtain the
AuxNLS hierarchy by varying with respect to $\psi,\bar{\psi}$.
Finally, although it will be of no relevance for the work in the rest of
this paper, I have indicated in figure 1 the first equations of
two more hierarchies in the NLS
tower, the well-known derivative NLS (DNLS) hierarchy [22], and a
quintic NLS (QNLS) hierarchy. Solutions of DNLS
can be obtained via Miura maps from both the PotNLS and AuxNLS
equations, and  generate solutions of NLS2. Solutions of QNLS can be
obtained via Miura maps from PotNLS, and generate solutions of both
NLS and NLS2. All details are in figure 1. The DNLS hierarchy
is obtained from the action by varying it with respect to the QNLS
variables and vice-versa.

Direct verification of everything in the above paragraph is mostly
tedious but straightforward. Amongst the harder manipulations is
finding the variation of the first term in (3.1) (which
I will call $S_0$) with respect to the sets of variables $\{A,B\}$,
$\{U,V\}$, $\{\psi,\bar{\psi}\}$, $\{P,Q\}$, $\{\tilde{P},
\tilde{Q}\}$, $\{p,q\}$. $S_0$ is a non-local functional of any
of these sets of variables. The formulae for the variation are:
$$\eqalign{
  \delta S_0&=\int dxdt~\left[ 2\left({{S_xT_t}\over{T_x}}-S_t\right)
                      \delta A ~+~ {{T_t}\over{T_x}}\delta B\right]  \cr
             &=\int dxdt~\left[ \left(\sigma_t+{{\tau_{tx}}\over{2\tau_x}}
                       \right)\delta U ~-~ {{\tau_{t}}\over{2\tau_x}}
                       \delta V\right] \cr
             &=\int dxdt~\left[ (q_t-q^2p_t)e^{-2h}\delta\psi ~+~
                       p_te^{2h}\delta\bar{\psi} \right]\cr
             &=\int dxdt~\left[ e^{2h} \tilde{Q}_t\delta P ~+~
                       e^{-2h}\tilde{P}_t\delta Q\right]\cr
             &=\int dxdt~\left[- e^{-2h}Q_t\delta\tilde{P} ~-~
                       e^{2h} P_t\delta\tilde{Q} \right]\cr
             &=\int dxdt~\left[ (e^{2\bar{h}}\psi_t-e^{2h}\bar{\psi}_t)
                     \delta p ~-~ e^{-2h}\psi_t\delta q\right]
          }\eqno{(3.4)}$$
For illustrative purposes I also give here a formula
(which also is quite intricate to obtain) for the
variation of the hamiltonian terms of (3.1) with respect to $\{p,q\}$
(this calculation is required to recover the usual NLS hierarchy from
(3.1)); if ${\cal F}$ is an arbitrary functional of $A,B$ we find
$$ \delta{\cal F} = \int dxdt~ \left[
        (\psi{\cal F}_A-(\bar{\psi}{\cal F}_B)_x)
        (e^{-2h}\delta q-e^{2\bar{h}}\delta p) ~+~
        (({\textstyle{1\over 2}}\psi^{-1}{\cal F}_{Ax}-\bar{\psi}{\cal F}_B)_x
         -\bar{\psi}{\cal F}_A)e^{2h}\delta p
                  \right]\eqno{(3.5)}$$
where ${\cal F}_A,{\cal F}_B$ denote functional derivatives of ${\cal F}$
with respect to $A,B$ respectively. Combining (3.5) and the appropriate part
of (3.4) it is clear how to vary $S_{NLS}$ with respect to $\{p,q\}$;
the usual form of the NLS hierarchy is recovered from the resulting
equations by using the results ${\cal F}_A=2\partial_x^{-1}(\bar{\psi}
{\cal F}_{\bar{\psi}}-\psi{\cal F}_{\psi})$, ${\cal F}_B=\psi^{-1}
{\cal F}_{\bar{\psi}}$.

The other important property of the action $S_{NLS}$
is that the canonical Poisson brackets associated with it [4] coincide
with the second hamiltonian structure of the NLS hierarchy (up to an
overall constant). These brackets are of course determined purely by
$S_0$, the only part of the action with a time derivative. Thus in the sequel
we will be quantizing the NLS hierarchy using its second hamiltonian structure
(in contrast to [13] where the first hamiltonian structure is used).

\vskip.2in

\noindent{\it 3b. The related conformal field theories}

Following [4] we would like to relate the term $S_0$ in $S_{NLS}$
to a conformal field theory. In fact {\it two} conformal
field theories emerge from $S_0$, essentially because
we can treat either the variables
$S,T$ or $\sigma,\tau$ as fundamental; the choice here determines the nature
of the fields $h,\bar{h}$ that appear in the free action $S_0$. If we
choose $S,T$ as fundamental fields, then $h,\bar{h}$ satisfy the constraint
${\cal H}_2=0$; if we choose $\sigma,\tau$ as fundamental fields,
then $h,\bar{h}$ satisfy the constraint ${\cal H}_1=0$. The two different
constraints give two very different theories; the well-defined fields in
each of the two theories have to commute with the relevant constraint,
so in the theory based on variables $S,T$ well-defined fields are
functionals of $A,B$, and in the theory based on variables $\sigma,\tau$
well-defined fields are functionals of $U,V$.

We can deduce at once that the theory based on variables
$\sigma,\tau$ is not going to be very interesting. Looking at (2.25) we
see that the variables $U,V$ decouple (at the classical level, but this
will remain true at the quantum level). The conformal field theory associated
with $S_0[\sigma,\tau]$,
for appropriate values of the coupling constant $\tilde{k}$,
consists of a $c=1$ scalar field and a minimal model. The significance of
the NLS hamiltonians in this theory will be that if we perturb the model
by adding a term proportional to a suitably quantized version of
$\int dt~{\cal H}_2$, then we expect the model to remain integrable,
with conserved quantities given by quantum analogs of the NLS hamiltonians.
But from the form of ${\cal H}_2$ it seems that such a
perturbation will be of very limited physical interest. So I will not
investigate this further.

In the theory based on variables $S,T$, however, we will use ${\cal H}_1$ as
our perturbation, and this looks more interesting. So we need to quantize
$S_0$ with ${\cal H}_2=0$ imposed as a constraint. Defining fields
$\phi_1,\phi_2$ by
$$ \eqalign{ h&={{-i}\over{2\sqrt{k+2}}}(\phi_1-\phi_2) \cr
             \bar{h}&={{-i}\over{2\sqrt{k+2}}}(\phi_1+\phi_2) }\eqno{(3.6)}$$
where $k$ is related to $\tilde{k}$ by $\tilde{k}=-(k+2)/2\pi$, we have
$$ S_0={1\over{8\pi}}\int dxdt~ (\phi_{1x}\phi_{1t}-\phi_{2x}\phi_{2t})
                    \eqno{(3.7)}$$
Operators in the theory depend on $x$, so we can translate them into operators
that are holomorphic in one complex variable $z$ in the standard
fashion [14] (identifying $x$ with the angular coordinate of the unit circle
in the $z$-plane). The normalization of (3.7) is such that two point
functions are given by
$$ \eqalign{ \langle\phi_1(z)\phi_1(w)\rangle&=-2\ln(z-w) \cr
             \langle\phi_2(z)\phi_2(w)\rangle&=+2\ln(z-w) }\eqno{(3.8)}$$
We wish to quantize (3.7) subject to a constraint
$$ \oint dz J(z) = 0 \eqno{(3.9)}$$
where we will take $J(z)=\phi_{2z}\exp{(i\phi_1/\sqrt{k+2})}$ (we can
apparently add a total derivative to $J$ without affecting the constraint;
this is simply the most compact choice).
Looking at [23] or [24], we see already that our theory
for integer $k$ has much in common
with the bosonized  ${\bf Z}_k$ parafermion theory;
both are expressed in terms of two bosons, one with negative signature,
and in the parafermion theory $\oint dz~J(z)$ is one of the screening
operators (so the states in the theory are characterized by the condition
that they are annihilated by $\oint dz~J(z)$, in a certain sense).
For $k<-2$ our theory has the features of the $SL(2,{\bf R})_{-k}/U(1)$
coset model [12]. The analogy continues:
classically, we had four composite fields $\psi,\bar{\psi},A,B$ all of
which commute with ${\cal H}_2$; $A$ has a quotient when expressed in terms
of $h,\bar{h}$, so its quantum analog will be difficult to handle,
but we would expect $\psi,\bar{\psi},B$ all to have sensible
quantum analogs (which commute with the constraint). Alas,
assuming we quantize the constraint
by a simple normal-ordering of the classical expression,
it is too much to expect that we might derive the quantum analogs of
$\psi,\bar{\psi},B$ by normal-ordering of the classical expressions.
We find we need to modify the classical expressions
$$ \eqalign{
  \psi&= {{-i}\over{2\sqrt{k+2}}}(\phi_{1x}-\phi_{2x})
       \exp\left({{i\phi_2}\over{\sqrt{k+2}}}\right) \cr
  \bar{\psi}&= {{-i}\over{2\sqrt{k+2}}}(\phi_{1x}+\phi_{2x})
        \exp\left({{-i\phi_2}\over{\sqrt{k+2}}}\right) \cr
  B&={1\over{4(k+2)}}(-\phi_{1x}^2 + \phi_{2x}^2) }\eqno{(3.10)}$$
to quantum expressions
$$ \eqalign{
 \psi&= :{{-i}\over{2\sqrt{k+2}}}
      \left(\sqrt{{k+2}\over{k}}\phi_{1x}-\phi_{2x}\right)
      \exp\left({{i\phi_2}\over{\sqrt{k}}}\right): \cr
 \bar{\psi}&=:{{-i}\over{2\sqrt{k+2}}}
      \left(\sqrt{{k+2}\over{k}}\phi_{1x}+\phi_{2x}\right)
      \exp\left({{-i\phi_2}\over{\sqrt{k}}}\right): \cr
  B&=:{1\over{4(k+2)}}
    \left(-\phi_{1x}^2 + \phi_{2x}^2+{{2i}\over{\sqrt{k+2}}}\phi_{1xx}\right):
          }\eqno{(3.11)}$$
which are (up to normalizations) the usual expressions for $\psi_1,
\psi_1^{\dagger},T$ in the bosonized ${\bf Z}_k$ parafermion model [23][24],
or $\psi_{\pm 1},T$ in the bosonized $SL(2,{\bf R})_{-k}/U(1)$ model [12].
Note that (up to overall normalization) the quantum
fields $\psi,\bar{\psi}$ are the {\it only} fields of the form
$:(\lambda\phi_{1x}+\mu\phi{2x})e^{i\nu\phi_2}:$ (with $\lambda,\mu,\nu$
constants) that commute with the constraint; similarly the improvement
we have to make to $B$ in the quantum theory is {\it uniquely} dictated.
So we have given here a novel approach to the construction of the bosonized
parafermionic theory; in the usual approaches [23][24] the discovery of
the screening operator $:J(z):$ is very ad hoc, whereas in our approach
it arises very naturally, as the one object transcribed directly from the
classical theory to the quantum theory, and used to determine the correct
quantized forms of all other operators of interest.

At this point it is appropriate to mention that there are features of the
${\bf Z}_k$ parafermion theory which it seems we cannot derive from
our classical theory, such as the other screening operators used in [24].
Therefore I have taken care not to call the $S_0$ an action for
the parafermion theory, rather I have just described it as {\it related} to
the parafermion theory (in the specific sense that both the operators in the
quantum theory based on $S_0$ and in the parafermion theory have to
commute with the constraint (3.9)).

What is the quantum analog of the NLS hamiltonians? In the classical theory
the NLS hamiltonians were quantities that commuted with both ${\cal H}_1$
and ${\cal H}_2$. So in the quantum theory, the NLS hamiltonians will have
to commute with not only the constraint, but also with another operator which
is a quantized version of ${\cal H}_1$. Here a problem appears; there
is apparently no natural way to specify the quantum analog of ${\cal H}_1$.
This actually will turn out to be to our advantage; naively quantizing
${\cal H}_1$, we would take its quantum analog to be
$$ \int dx~:\exp\left({{-i\phi_1}\over{\sqrt{k+2}}}\right):   \eqno{(3.12)}$$
but we will allow ourselves the freedom of generalizing this to
$$ \int dx~:\exp\left({{-il\phi_1}\over{\sqrt{k+2}}}\right):   \eqno{(3.13)}$$
where $l$ is a constant. Allowing this we see at once that {\it the quantum
NLS hamiltonians are the conserved quantities under thermal perturbations of
the parafermion models} (and similar perturbations of the $SL(2,{\bf R})/U(1)$
coset models). In the parafermion theory the significance
of requiring operators to commute with the operator (3.12) is that they
are conserved quantities under perturbation by the first thermal operator [11];
requiring commutation with the operator (3.13) (for positive integer $l$)
gives us conserved quantities under perturbation by the $l$th thermal operator.
So we see that by allowing the possibility (3.13) we stand the chance to
learn more. We will return to the explicit construction of the
first few quantum NLS hamiltonians in section 3d; we will see that
for $l\not=1$ not all the NLS hamiltonians survive the quantization, but
for $l=1$ it is possible they do.

\vskip.2in

\noindent{\it 3c. Understanding $S_0$}

\noindent
Although the form of $S_0$ given in section 3a is obvious to guess, given
that we know we want the $j,\bar{j}$ Poisson bracket structure (2.17b),
$S_0$ has a much deeper origin, which in particular will explain the formula
for $\Omega_2$ in (2.46). Let $g(x,t)$ be an $SL(2)$ valued
function, with a Gauss decomposition as in (2.43), and with the
diagonal component of $g^{-1}g_x$ vanishing, so (2.44) holds. Then
$S_0$ (as a functional of $S,T$, which is how we are now treating it)
is simply the WZW action for $g$. This can
be checked using the formulae of [25] (note that the crucial
result we have used, that we need to supplement the action $S_0$ as a
functional of $h,\bar{h}$ with a constraint, was also obtained in [25]).
The formula for $\Omega_2$ in (2.46) is  the  symplectic
form associated with the WZW action, pulled back to the space of
$SL(2)$ valued functions $g(x)$ with the diagonal component of $g^{-1}g_x$
vanishing. Note that in general if we specify the hamiltonian structure of
some model by giving a symplectic form, it is very easy to restrict the
model by some constraint; we simply pull back the form. However, if the
hamiltonian structure is specified via a set of Poisson brackets, we have
to do  a Dirac reduction; in  our case we have to reduce the $SL(2)$ Kac-Moody
algebra, the Poisson bracket algebra of the $SL(2)$ WZW model, by the
constraint that the diagonal current vanishes. This gives the original
second Poisson bracket structure of the NLS equation, (2.3). We see that
{}from (2.3) we might have directly deduced that an appropriate action for the
NLS hierarchy would be the SL(2) WZW model for $g$ with the diagonal component
of $g^{-1}g_x$ vanishing, plus a sum of the NLS hamiltonians (the WZW action,
being first order in (light-cone) time derivatives, gives no hamiltonian,
but determines Poisson brackets).

For certain $G/H$ coset models, Park [26] has proposed considering an
action $S_{WZW}(m_g)$. This action is a functional of
a $G$ valued function $g$, and $m_g$ is defined so that $m_g^{-1}m_{gx}$ is
orthogonal to the Lie algebra of $H$ in the Lie algebra of $G$.
Our $S_0$ is an instance of Park's action. By a simple path integral
manipulation, Park has shown the equivalence of the theory based on his action
with the theory based on the usual gauged WZW action for coset models [27].
But as I have said above, some features of the coset model are not
immediately apparent from the canonical quantization of the Park action.
Clarification of this issue would be interesting.

In addition to the relation to the Park action, it would seem that $S_0$
also has a direct relation to the usual gauged WZW action. Bardakci et al.
[28] have shown that in the classical $G/H$ gauged WZW model there are certain
composite fields that satisfy a ``classical parafermion algebra'', i.e.
the Dirac reduction of the $G$ Kac-Moody algebra by the constraints that the
$H$ currents are set to zero (of course, in the theory defined by Park's
action it is completely clear that there are fields satisfying this algebra).
This being the case, we should be able to identify $S_0$ as (maybe a term in)
a gauged WZW model. This may help to find the additional information
needed for a complete Lagrangian description for the parafermion models;
however we will not investigate this here.

\vskip.2in

\noindent{\it 3d. The first few quantum NLS hamiltonians}

\noindent We now return to the explicit construction of the first few
quantum NLS hamiltonians. The spin $s-1$ quantum NLS hamiltonian should
have the form $\int dx~:T_s:$ where $T_s$ is some spin $s$ operator
constructed out of $\phi_{1x},\phi_{2x}$ and their derivatives. Clearly
we can freely add to $T_s$ a total derivative, and we also would hope
that the quantum NLS hamiltonians share the symmetry of the classical
NLS hamiltonians that under $\phi_2\rightarrow -\phi_2$, $T_s\rightarrow
(-1)^sT_s$. The overall normalization of $T_s$ is unimportant for our
purposes but we will nevertheless choose not to fix it. Given all
this, we should look for the first few $T$'s in the form
$$ \eqalign{ T_2&=\alpha_1\phi_{1x}^2 + \alpha_2\phi_{2x}^2 \cr
             T_3&=\beta_1\phi_{1x}^2\phi_{2x} + \beta_2\phi_{2x}^3 +
                  \beta_3\phi_{1x}\phi_{2xx} \cr
             T_4&=\gamma_1\phi_{1x}^4 + \gamma_2\phi_{1x}^2\phi_{2x}^2
                 +\gamma_3\phi_{22}^4 + \gamma_4\phi_{1x}\phi_{2x}\phi_{2xx}
                 +\gamma_5\phi_{1xx}^2 + \gamma_6\phi_{2xx}^2 }
            \eqno{(3.14)}$$
We require the quantum NLS hamiltonians to commute with the quantum analogs
of ${\cal H}_1$ (which depends on the parameter $l$) and ${\cal H}_2$.
Defining
$$ \eqalign{\Phi_1 &= \phi_{2x}e^{i\mu\phi} \cr
            \Phi_2 &= e^{-il\mu\phi} }  \eqno{(3.15)}$$
where we have written $\mu=1/\sqrt{k+2}$, this means that we need to enforce
the following OPEs:
$$ T_s(z')\Phi_i(z) = \sum_{r>0} {{\Psi_{si,r}(z)}\over{(z'-z)^{r+1}}}
                    + {{\partial_z\Psi_{si,0}(z)}\over{z'-z}} + {\rm reg.}
            \eqno{(3.16)}$$
for $i=1,2$. Here $\Psi_{si,r}$ denote some normal ordered operators
whose form is unimportant for us.
In general what seems to happen is that the OPE of $T_s$
with $\Phi_1$ determines some relations between the coefficients of $T_s$,
and the OPE with $\Phi_2$ fixes the remaining coefficients (up to an overall
normalization), and also gives constraints on $l$. The $T_2$ case is
somewhat trivial; we just find we need $\alpha_2=-\alpha_1$, and any $l$ is
permitted. For $T_3$ from the $\Phi_1$ OPE we obtain
$$ \eqalign{\beta_2 &= \left({{2\mu^2}\over 3}-1\right)\beta_1 \cr
            \beta_3 &= {2\over {i\mu}}(\mu^2-1) \beta_1  }\eqno{(3.17)}$$
The $\Phi_2$ OPE yields the requirement $l=1$. So only if we quantize
${\cal H}_1$ in the naive fashion, as explained in section 3b, do we
obtain a spin 2 conserved quantity. Moving on to $T_4$ we see some of the
features I expect to be present in the general case. We find from the
$\Phi_1$ OPE:
$$ \eqalign{ \gamma_3&=(2\mu^2-1)\gamma_1 + (\mu^2-1)\gamma_2 \cr
             \gamma_4&={2\over{i\mu}}\left(
                      6(2\mu^2-1)\gamma_1 + (4\mu^2-3)\gamma_2\right) \cr
             \gamma_5&=2\left({1\over{\mu^2}}-2\mu^2\right)\gamma_1
                       +\left({1\over{\mu^2}}-2\right)\gamma_2   \cr
             \gamma_6&=2\left(2\mu^2-1-{1\over{\mu^2}}\right)\gamma_1
                       +\left({{2\mu^2}\over 3}+1-{1\over{\mu^2}}\right)
                                           \gamma_2   }
                    \eqno{(3.18)}    $$
Thus one coefficient in $T_4$ is not fixed. The $\Phi_2$ OPE gives
$$ \eqalign{ \gamma_4&={4\over{il\mu}}(l^2\mu^2-1)\gamma_2 \cr
             \gamma_5&=4\left(3-l^2\mu^2-{1\over{l^2\mu^2}}\right)\gamma_1
           }  \eqno{(3.19)}$$
These fix the last coefficient and give a consistency condition for $l$,
which is solved to give $l=1$ or $l=2$. Note the coefficients in $T_4$
depend on the choice of $l$; for example for $l=1$ we find $\gamma_2=
-6\gamma_1$ and for $l=2$ we find $\gamma_2=3(2\mu^2-1)\gamma_1$.

The scheme of
results above is in agreement with the literature. For the ${\bf Z}_k$
parafermion model, spin 2 and 3 conserved quantities in the first thermal
perturbation, and a spin 3 conserved quantity in the second thermal
perturbation were given in the appendix of [11]. For the $SL(2,{\bf R})/U(1)$
coset model spin 2 and spin 3 conserved quantities, corresponding to
our results for $l=1$, were given in  [12]. The
simple forms of the coefficients $\gamma_3,...,\gamma_6$ (after
setting $\gamma_2=-6\gamma_1$) seem quite remarkable when compared to the
complexity of equations (5.7),(A.1),(A.2),(B.1) in [12], but explicit checks
of $\gamma_4$ and $\gamma_5$ (which are the two formulae needed to determine
$l$) show the equivalence. We have avoided
many of the steps of [12], and indeed lost some information found
there, but from the point of view of explicit construction of the charges
the method presented here is more efficient.

One piece of information we have not obtained is the
fact that there is a $W-$algebra in all of our models, a $W_k$
algebra for the ${\bf Z}_k$ parafermions and a $\hat{W}_{\infty}(-k)$ algebra
for the $SL(2,{\bf R})_{-k}/U(1)$ models [12]. From [11] we know that for
the $l=1$ deformation of the ${\bf Z}_k$ parafermion model there should be
conserved quantities for all spins which are not zero modulo $k$. The
spin 2 and 3 conserved quantities constructed above seem to exist even
for $k=2$ and $k=3$. They must vanish by the correct choice of certain
$k$-dependent normalization factors, but I cannot at present see a method of
fixing the normalizations. The truncation of the parafermionic $W-$algebra
has been studied in depth in [29], from which it seems that the
reason the $W-$currents of spin $\ge k$ can be ignored in the ${\bf Z}_k$
parafermion model is also a result of correct normalization in their
definition, but it is not obvious how the normalization should be chosen.
This question needs further study. But for now we have to accept the
results from above for the parafermion case with a little caution.
Some constraint on the normalization of the quantum NLS hamiltonians
is obtained by requiring the correct classical limit, but this is
not enough to resolve the question raised here. The classical limit
is obtained by writing everything in terms of $h,\bar{h}$ via (3.6)
and taking $k\rightarrow\infty$; the normalizations of the classical NLS
hamiltonians can be related by the recursion relation (2.20).

For the coset model case there is apparently no problem of the type
mentioned above (except possibly for certain rational values of $k$).
The existence of a $W_{\infty}$ type algebra in the $SL(2,{\bf R})/U(1)$ coset
model seems in fact to be of very limited physical significance (just as
the fact that one can build operators satisfying a $W_{\infty}$ type algebra
{}from a single boson is only really a result of interest from the point
of view of $W_{\infty}$ algebra representation theory). We have seen how to
construct the charges of the theory without mentioning the $W-$algebra,
and we have also one interesting new piece of information from our approach:
the conserved charges of the $SL(2,{\bf R})/U(1)$ coset model studied in
[12] (see also [30]) are only one particular choice for the conserved
charges of the model, those appropriate to the first thermal perturbation.
There are other infinite sets as well, appropriate to other perturbations.
Whether any of these sets have physical significance, as suggested in [12],
remains to be seen.

\vskip.2in

\noindent{\it 4.Concluding Remarks}

\noindent It seems reasonable to suspect that the patterns we have seen in
section 3d will continue and that we will find a complete set of
quantum NLS hamiltonians for $l=1$; computer algebra could easily be
used to check the constraints on $l$ for the next few conserved quantities
to exist. It also seems reasonable to guess that  quantum hamiltonians
of the generalized NLS equations of Fordy and Kulish [31] should exist,
and give conserved quantities in deformed $G/H$ coset models where $G/H$
is a homogeneous or hermitian symmetric space. These models are
of some interest (and particularly their supersymmetric generalizations)
so this might be an worthwhile problem to investigate.

\vskip.2in

\noindent{\bf Acknowledgements}

\noindent Useful discussions with Lisa Jeffrey, Boris
Kupershmidt, Olaf Lechtenfeld, Ed Witten and
particularly Didier Depireux are gratefully acknowledged.
I also thank Didier Depireux and Pierre Mathieu for some comments on
the manuscript. This work was supported by a
grant in aid from the U.S.Department of Energy, \#DE-FG02-90ER40542.

\vskip.2in

\noindent{\bf References}

\noindent
\item{[1]} The original observation of this type is due to
           J.-L. Gervais, {\it Phys.Lett.B} {\bf 160} (1985) 277.
\item{[2]} A.B.Zamolodchikov, {\it Adv.Stud.in Pure Math.} {\bf 19} (1989) 641.
\item{[3]} P.Mathieu, {\it Nucl.Phys.B} {\bf 336} (1990) 338.
\item{[4]} J.Schiff, {\it The KdV Action and Deformed Minimal Models},
      Institute for Advanced Study Preprint IASSNS-HEP-92/28 (revised version).
\item{[5]} L.A.Dickey, {\it Ann.N.Y.Acad.Sci.} 410 (1983) 301.
\item{[6]} G.Wilson, {\it  Phys.Lett.A} {\bf 132} (1988) 45.
\item{[7]} G.Wilson, {\it Quart.J.Math.Oxford} {\bf 42} (1991) 227,
      {\it Nonlinearity} {\bf 5} (1992) 109, and in {\it Hamiltonian
     Systems, Transformation Groups and Spectral Transform Methods},
     ed. J.Harnad and J.E.Marsden, CRM (1990).
\item{[8]} T.J.Hollowood, L.Miramontes, A.Pasquinucci and C.Nappi, {\it
    Nucl.Phys.B} {\bf 373} (1992) 247; C.S.Xiong, talk delivered at
    the CAP/NSERC workshop on Quantum Groups, Integrable Models and
    Statistical Systems, Kingston, Ontario, Canada (1992).
\item{[9]} V.G.Drinfeld and V.V.Sokolov, {\it Jour.Sov.Math.} {\bf 30}
            (1985) 1975.
\item{[10]} M.F.de Groot, T.J.Hollowood and J.L.Miramontes,
     {\it Comm.Math.Phys.} {\bf 145} (1992) 57.
\item{[11]} V.A.Fateev, {\it Int.J.Mod.Phys.A} {\bf 6} (1991) 2109.
\item{[12]} I.Bakas and E.Kiritsis, {\it Beyond the Large N limit: Non-linear
     $\rm W_{\infty}$ as Symmetry of the $\rm SL(2,R)/U(1)$ Coset Model},
     University of California/Berkeley/Maryland preprint
    UCB-PTH-91/44, LBL-31213, UMD-PP-92-37.
\item{[13]} M.Omote, M.Sakagami, R.Sasaki and  I.Yamanaka, {\it Phys.Rev.D}
      {\bf 35} (1987) 2423.
\item{[14]}  R.Sasaki and I.Yamanaka, {\it Comm.Math.Phys.} {\bf 108}
    (1987) 691, {\it Adv.Stud.in Pure Math.} {\bf 16} (1988) 271. See
   also  B.A.Kupershmidt and P.Mathieu, {\it Phys.Lett.B} {\bf 227} (1989) 245.
\item{[15]} F.Magri, {\it J.Math.Phys.} {\bf 19} (1978) 1156.
\item{[16]} N.J.Burroughs, M.F.DeGroot, T.J.Hollowood and J.L.Miramontes,
    {\it Generalized Drinfel'd-Sokolov Hierarchies II: The Hamiltonian
     Structures}, Princeton University/Institute for Advanced Study preprint
     PUPT-1263, IASSNS-HEP-91/42.
\item{[17]} P.A.M.Dirac, {\it Lectures on Quantum Mechanics},
      Yeshiva University (1964).
\item{[18]} J.Schiff, {\it Self-Dual Yang-Mills and the Hamiltonian
          Structures of Integrable Systems}, in preparation.
\item{[19]} B.A.Kupershmidt, {\it Comm.Math.Phys.} {\bf 99} (1985) 51.
\item{[20]} D.A.Depireux, {\it Mod.Phys.Lett.A} {\bf 7} (1992) 1825.
\item{[21]} P.J.Olver {\it Applications of Lie Groups to Differential
    Equations}, Springer-Verlag (1986).
\item{[22]} D.J.Kaup and A.C.Newell, {\it J.Math.Phys} {\bf 19} (1978) 798.
\item{[23]} D.Nemeschansky, {\it Phys.Lett.B} {\bf 224} (1989) 121;
    {\it Nucl.Phys.B} {\bf 363} (1991) 665.
\item{[24]} T.Jayaraman, K.S.Narain and M.H.Sarmadi, {\it Nucl.Phys.B}
    {\bf 343} (1990) 418.
\item{[25]} A.Gerasimov, A.Morozov, M.Olshanetskii, A.Marshakov and
        S.Shatashvili, {\it Int.J.Mod.Phys.A} {\bf 5} (1990) 2495.
\item{[26]} Q-Han Park, {\it Phys.Lett.B} {\bf 223} (1989) 175.
\item{[27]} See for instance D.Karabali and H.J.Schnitzer, {\it Phys.Lett.B}
   {\bf 216} (1989) 307; {\it Nucl.Phys.B} {\bf 329} (1990) 649.
\item{[28]} K.Bardakci, M.Crescimanno and E.Rabinovici, {\it Nucl.Phys.B}
    {\bf 344} (1990) 344; K.Bardakci, M.Crescimanno and S.A.Hotes,
    {\it Nucl.Phys.B} {\bf 349} (1991) 439.
\item{[29]} F.J.Narganes-Quijano, {\it Int.J.Mod.Phys.A} {\bf 6} (1991) 2611.
\item{[30]} F.Yu and Y.-S.Wu, {\it Nucl.Phys.B} {\bf 373} (1992) 713;
   {\it An Infinite Number of Commuting $\hat{W}_{\infty}$ Charges in the
   SL(2,R)/U(1) Coset Model}, Utah preprint  UU-HEP-92/11.
\item{[31]} A.P.Fordy and P.P Kulish, {\it Comm.Math.Phys.} {\bf 89} (1983)
     427.

\bye